\documentclass{article}

\usepackage{arxiv}
\usepackage[T1]{fontenc}
\usepackage[utf8]{inputenc}
\usepackage{amsmath}
\usepackage{booktabs}
\usepackage{tabularx}
\usepackage{longtable}
\usepackage{array}
\usepackage{calc}
\usepackage[table]{xcolor}
\usepackage{modernpaper}
\usepackage{textcomp}
\usepackage{listings}
\usepackage{pgfplots}
\pgfplotsset{compat=1.18}
\usepackage[numbers]{natbib}
\usepackage{url}
\makeatletter
\g@addto@macro{\UrlBreaks}{\do\a\do\b\do\c\do\d\do\e\do\f\do\g\do\h\do\i\do\j\do\k\do\l\do\m\do\n\do\o\do\p\do\q\do\r\do\s\do\t\do\u\do\v\do\w\do\x\do\y\do\z\do\A\do\B\do\C\do\D\do\E\do\F\do\G\do\H\do\I\do\J\do\K\do\L\do\M\do\N\do\O\do\P\do\Q\do\R\do\S\do\T\do\U\do\V\do\W\do\X\do\Y\do\Z\do\0\do\1\do\2\do\3\do\4\do\5\do\6\do\7\do\8\do\9\do\-}
\makeatother
\usepackage{hyperref}

\providecommand{\tightlist}{\setlength{\itemsep}{0pt}\setlength{\parskip}{0pt}}
\title{Look Here, Then Decide: Small Decision Models in a Harness Find Memory-Safety Vulnerabilities at Scale on Commodity GPUs}
\author{{\setlength{\tabcolsep}{2pt}\begin{tabular*}{\textwidth}{@{\extracolsep{\fill}}cccc@{}}
\textbf{Tim De Wachter}\textsuperscript{1,2} & \textbf{Kyryl Volkov}\textsuperscript{1,2} & \textbf{Myrto Eleftheria Gkogkou}\textsuperscript{2} & \textbf{Mathieu Huysman}\textsuperscript{1,2}\\[-4pt]
{\fontsize{8.5}{10}\selectfont\texttt{tim@deusrex.ai}} & {\fontsize{8.5}{10}\selectfont\texttt{kyryl.volkov@deusrex.ai}} & {\fontsize{8.5}{10}\selectfont\texttt{myrto.gkogkou@deusrex.ai}} & {\fontsize{8.5}{10}\selectfont\texttt{mathieu@deusrex.ai}}
\end{tabular*}}\\[8pt]
\textsuperscript{1} Cyrex \quad \textsuperscript{2} Deus Rex}
\date{}

\renewcommand{\shorttitle}{Small Decision Models for Vulnerability Discovery}
\renewcommand{\headeright}{}
\renewcommand{\undertitle}{}
\hypersetup{
  colorlinks=true,
  citecolor=PaperAccent,
  linkcolor=PaperInk,
  urlcolor=PaperAccent,
  pdftitle={Look Here, Then Decide: Small Decision Models in a Harness Find Memory-Safety Vulnerabilities at Scale on Commodity GPUs},
  pdfauthor={Tim De Wachter, Kyryl Volkov, Myrto Eleftheria Gkogkou, Mathieu Huysman}
}

\begin{document}

\maketitle

\begin{abstract}
We ask whether small, open-weight \emph{decision models} can support vulnerability discovery in open-source software at scale and what the surrounding system must provide. These models answer a typed question in one forward pass, returning scores for the permitted answers rather than text. Our harness selects widely packaged C and C++ projects and uses deterministic incomplete-fix miners and shape localizers to decide \emph{where to look}. Intern-Decision-4B, a 4B-parameter open-weight decision model, then scores every candidate function. The scores \emph{rank} the queues rather than accept or reject candidates, apart from a low reach-score floor that some queues kept. The search uses no frontier model.

Alone, the model is weak: it separates vulnerable from benign functions with AUC 0.73, but a vulnerable function from its own patch with only 0.56. The model is more useful for ordering candidates. On 75 public CVEs from projects without any of our findings, the fixed function reached its repository's top 25 in 21\% of cases---1.4 times as often as under random order of the same queues. Recall is bounded mainly by where the localizers look. On the campaign's own queues the model's order placed 7.5 times more confirmed findings in the top 10 than random order, an observed, selection-biased upper bound, because those queues were chosen for containing a confirmed finding and were reviewed in the model's order. On three desktop-class GB10 systems, the harness searched more than 4,000 projects, mining them at 900--1,100 repositories per hour across two of the systems. The campaign confirmed 187 memory-safety findings; by worst-case CVSS v3.1, 12 are critical, 148 high, 23 medium and 4 low. The decision step runs at 4.4 decisions per second on one GB10 and at 9.8 on a consumer RTX 4090 gaming PC, with matching scores. For this model and harness, a small decision model supports vulnerability discovery through prioritization, while deterministic analysis and verification remain essential. All confirmed findings are being disclosed to maintainers under coordinated disclosure before publication.
\end{abstract}


\section{Introduction}\label{sec:intro}

C and C++ underpin operating systems, libraries and network services used by millions of people. Much of this code handles untrusted files or network data, and a mistake in a memory operation can become a security vulnerability. Memory corruption has long persisted in C and C++ software \citep{szekeres2013eternal}, and memory-safety defects remain the largest class of serious vulnerabilities in such code \citep{thomas2019msrc70, chromium_memsafety}. A defect in a widely packaged library can affect many downstream applications. Finding these defects early gives maintainers time to coordinate fixes.

Static application security testing (SAST) is one of the most widely used approaches. Rule engines such as Semgrep, its open fork Opengrep, and CodeQL \citep{semgrep, opengrep2025, codeql, avgustinov2016ql} inspect source code without running it. They use syntax trees and data-flow analysis to find code that matches a known-bad pattern. These tools are fast and explainable, but do not decide whether the flagged code is reachable or exploitable. Empirical studies also report high noise and large recall gaps on real C vulnerabilities \citep{lipp2022staticc, charoenwet2024sastreview}.

Large language models (LLMs) offer another route to vulnerability discovery. Recent agentic systems combine language models with execution environments and analysis tools to investigate and reproduce suspected vulnerabilities \citep{glazunov2024naptime, bigsleep2024sqlite}. Anthropic later introduced Claude Mythos Preview, a frontier model available to selected partners, and reported its ability to identify high-severity vulnerabilities in widely used software \citep{anthropic2026glasswing, carlini2026mythospreview}. OpenAI reported that a single run of a cyber-specialised frontier model validated and wrote patches for 948 local privilege-escalation vulnerabilities in the Linux kernel, at a cost it gives as US\$1,213,430 in API credits, about US\$1,280 per vulnerability \citep{openai2026paths, openai2026gpt6cyber}. These developments show what LLM-based discovery can do. Applying frontier models to thousands of repositories, however, is expensive when each investigation needs many model calls and tool executions; access to some of the most capable models also remains restricted. Large-scale discovery is therefore hard for teams with limited resources.

\subsection{Motivation}\label{sec:motivation}

How can language models be used across large codebases without an expensive, in-depth analysis of every candidate? Running an agent on every function of thousands of repositories would need substantial compute, and most of the functions it read would contain no vulnerability. Deterministic analysis can narrow the search before a small model decides which candidates to investigate first.

In 2026 a model category suited to the second step appeared: the \emph{decision model}, or \emph{System-One} model, a category TypeSafe AI introduced with its Jev model \citep{typesafe2026jev}. Instead of producing an explanation, a decision model evaluates predefined questions and returns a score for each permitted answer in a single forward pass (Section~\ref{sec:background}). Each decision costs one prefill and no decoding. Its output orders a queue directly, with nothing to parse.

We study Intern-Decision-4B, a small open-weight decision model \citep{internlm2026interndecision}, as one component of a larger harness rather than as a standalone detector. Our central research question is:

\begin{quote}
\emph{Can small, open-weight decision models improve the prioritization of potential memory-safety vulnerabilities at scale, even when their standalone classification performance is limited? And what must the harness around them do?}
\end{quote}

\subsection{Threat model}\label{sec:threat}

We take the defender's view. Consider a team with a few desktop-class GPU systems, open-weight models and commodity build infrastructure, but no access to frontier models restricted to selected partners. Can it find memory-safety defects in widely packaged C and C++ software early, at a rate that helps maintainers fix them first? Our results suggest it can, for this model and harness. We read them as showing that the design of the harness matters more than the size of the model; that is an interpretation, not a controlled comparison of model sizes. The ranking ablation and recall study (Sections~\ref{sec:harness} and \ref{sec:recall}) measure parts of this interpretation; the evaluation does not isolate every component. All confirmed findings are being disclosed to maintainers under coordinated disclosure before publication (Section~\ref{sec:disclosure}).

\subsection{Contributions}\label{sec:contrib}

Our harness combines deterministic candidate generation, model-based ranking and subsequent verification. It selects projects by deployment reach and existing fuzzing coverage, mines their revision histories for incomplete security fixes, and searches for code shapes associated with unsafe memory operations. For example, a security patch may add a bounds check before a copy in one function while leaving a similar copy elsewhere unchanged; that sibling is a candidate, though not necessarily a vulnerability. Intern-Decision-4B scores every candidate. Stronger coding agents and human reviewers then verify the candidates they read. Candidate generation decides where to look, the decision model decides what to read first, and verification decides whether a vulnerability exists. Our contributions are:

\begin{enumerate}
\def\labelenumi{\arabic{enumi}.}
\tightlist
\item
  \textbf{An evaluation of small decision models as vulnerability classifiers.} On real CVE fixes, Intern-Decision-4B and six configuration baselines built from other small open classifiers separate vulnerable code from benign code but barely from its own patch (Section~\ref{sec:model}).
\item
  \textbf{The harness that makes a weak model useful.} Deterministic localizers and miners, a reach-ranked and under-fuzzed corpus, and ranking in place of thresholds. In localizer pilots, 30--33\% of the top 30 of a ranked queue were real candidates, against 6.7\% of a random sample; this is a yield observation by configuration, not an isolated ranking effect (Sections~\ref{sec:method} and \ref{sec:harness}).
\item
  \textbf{Measured ranking lift and recall.} We measure recall on 75 randomly drawn public CVEs from projects without any of our findings: ranking placed fixed functions within the review budget 1.3--1.4 times as often as random order of the same queues. An ablation on the campaign's real queues found 7.5 times as many confirmed findings in the top 10 as random order, an observed, selection-biased upper bound; we also publish the CVE selection rule and list (Sections~\ref{sec:harness} and \ref{sec:recall}).
\item
  \textbf{Speed on commodity hardware.} 4.4 decisions per second on one GB10 and 9.8 on a consumer RTX 4090 gaming PC under vLLM, with matching scores on both and on a laptop GPU, and a cost-and-shape comparison with Opengrep (Section~\ref{sec:speed}).
\item
  \textbf{A large-scale campaign.} More than 4,000 open-source C and C++ projects searched on three desk-side GPU systems, with mining at 900--1,100 repositories per hour: 187 confirmed memory-safety findings, with their worst-case CVSS v3.1 severity and every false-positive measure the record supports (Section~\ref{sec:findings}).
\end{enumerate}

\section{Background}\label{sec:background}

\subsection{Decision models and System-One-style inference}\label{sec:bgdecision}

We use \emph{decision model} for a language model used as a classifier: it answers a predefined question in one forward pass, without generating text. The term and the model category were introduced by TypeSafe AI with Jev \citep{typesafe2026jev}. Each question specifies an answer type, such as yes/no or a choice among labelled options. We read each label's score from the model's answer-token probabilities. A restricted softmax over those tokens, optionally temperature-scaled, gives a distribution over the labels. Although the scores sum to one, they are not necessarily calibrated probabilities of the outcomes being assessed (Section~\ref{sec:bgranking}).

The idea relates to \emph{P(True)}-style self-evaluation \citep{kadavath2022know} and to temperature scaling \citep{guo2017calibration}. It differs from LLM-as-judge setups that generate a rationale and a verdict \citep{zheng2023judging}, and from encoder classifier heads trained on vulnerability datasets \citep{feng2020codebert, guo2021graphcodebert, guo2022unixcoder, fu2022linevul, zhou2019devign}.

Several open-weight models are published for this use, all under Apache-2.0: Intern-Decision-4B (4.5B, fine-tuned by InternLM from Qwen3.5-4B) \citep{internlm2026interndecision, qwen2026qwen354b}; clef (27B) \citep{cloudflare2026clef}; pplx-decider-v1.1 (26B) \citep{perplexity2026pplxdecider}; and decider-chat-gemma4-31b (31B) \citep{mapika2026decider}. Intern-Decision-4B is not Jev but an independent open-weight member of the same class: its card calls it a ``structured decision model'', its responses follow the Jev envelope, and it reports a mean of 44 ms per query against 110 ms for Jev on the same benchmark \citep{internlm2026interndecision}. AnyJev turns any causal language model served by vLLM into such a decider by reading label log-probabilities \citep{zhang2026anyjev, kwon2023vllm}.

Two details matter in practice.

\begin{itemize}
\tightlist
\item
  \textbf{Read labels by id.} In one case a 31B model at temperature 1 put the ``No'' label below the top 20 tokens, so a fixed top-K read could not see it.
\item
  \textbf{Scores are not calibrated at the real base rate.} In one evaluation a model's mean stated probability was 0.35 against a true rate of 0.005. Scores must rank, not decide.
\end{itemize}

\subsection{Signals for vulnerability discovery}\label{sec:bgsignals}

Revision history and current code both offer clues about potential vulnerabilities. Previous security fixes are one source of evidence: a patch may add a necessary check at one location while leaving a similar operation elsewhere unchanged, or add a check that fails to exclude every unsafe value. These situations motivate patch-based analysis and research on recurring vulnerabilities \citep{xiao2020mvp, feng2024fire, liu2025incompletelinux}. If a patch adds a length check to one function that copies input into a fixed-size buffer, a similar copy in another function may warrant investigation. The missing check marks a location worth analysing, although callers may constrain the second function's inputs; it does not by itself establish a vulnerability. We call a tool that finds such candidates in revision history an \emph{incomplete-fix miner}.

Code patterns can also reveal candidates without a previous fix: input-dependent lengths used in buffer writes, arithmetic that can wrap before an allocation or copy, or externally supplied sizes passed to memory operations. We call a detector for one such pattern a \emph{shape localizer}. Both tools propose locations. Establishing a vulnerability requires examining the surrounding constraints and execution paths.

Fuzzing coverage also helps us choose projects. Fuzzing runs software on generated or mutated inputs to expose failures, and memory-error detectors such as AddressSanitizer flag invalid accesses during those runs \citep{serebryany2012asan}. Projects without known fuzzing coverage are reasonable search targets, although the absence of documented coverage does not establish that a project has never been fuzzed, and existing coverage does not establish the absence of vulnerabilities.

\subsection{Ranking, classification and calibration}\label{sec:bgranking}

A score can be used to classify candidates or to order them. Classification applies a threshold: candidates above it are labelled positive and those below negative. Ranking orders candidates by score so that higher-scoring ones are examined first. The two uses impose different requirements. A model may assign higher scores to promising candidates without supporting a reliable threshold for deciding whether any single function is vulnerable. Under a limited review budget, a useful ordering increases the number of real findings among the candidates read first; it does not establish the validity of any individual finding.

Calibration concerns whether predicted probabilities agree with observed frequencies \citep{guo2017calibration}: among candidates assigned a probability near 0.8, a calibrated predictor is right about 80\% of the time. Answer-token probabilities do not automatically have this property for vulnerability detection. The \emph{base rate} is the share of genuine vulnerabilities in the population examined; when it is small, even a modest false-positive rate produces many more false alarms than real findings. Evaluation must therefore consider the candidate population and the review budget, not only the model's ability to separate positive from negative examples. Our harness therefore uses decision-model scores to order the work and relies on contextual analysis and reproduction to confirm findings.

\section{Related work}\label{sec:related}

\textbf{Decision models for security (closest work).} The closest prior work is CT0, a cybersecurity model by Mayoral-Vilches \citep{mayoralvilches2026ct0}. CT0-Micro, at roughly 4B parameters, pairs a ``System 1'' fast path that answers typed security decisions in a single forward pass without generating tokens with a ``System 2'' tool-using agent mode behind a confidence gate. It reports 21.1\% of Cybench subtasks solved on the agentic path (against 7.6\% for its base model), IOC file screening at about 80 files per second, jailbreak detection at AUROC 0.966, and vulnerability-fact extraction that beats either mode alone at about a fifth of the generative cost. CT0 shows that System-One-style decision models work for security classification and triage. Our work complements CT0 by applying an off-the-shelf model of this class to vulnerability \emph{discovery} in real open-source code at scale. Our harness decides where to look and uses scores mainly to rank, keeping only a low reach-score floor on some queues; we report measured throughput and confirmed findings.

Theori's AIxCC system RoboDuck is the closest prior \emph{design}: it ranked static-analysis candidates by the log-probabilities of a single ``likely/unlikely'' LLM classification and sent only the top fifth to a costly agent step \citep{xint2025roboduck}. Needles at Scale likewise uses a low-cost language model to rank structurally pre-filtered Windows binary functions for review \citep{bommarito2026needles}. Two results support reading labels from one pass rather than from generated text: in SOC alert triage, chain-of-thought reasoning degraded the label-token probabilities that automated triage relies on \citep{khanna2026cybersecurity}, and hidden states separate vulnerable from secure code better than prompting does \citep{yu2025mixture}.

\textbf{Static analysis with learned triage.} Z-Ranking proposed ranking a static analyser's reports rather than filtering them \citep{kremenek2003zranking}. Heckman and Williams survey actionable-alert identification \citep{heckman2011sla}; Kang et al.~show how leakage and duplication inflated earlier learned results \citep{kang2022falsealarms}. LLMs now inspect static bug warnings \citep{wen2024llm4sa}, cut false positives in industrial static bug detection \citep{du2026industryfp}, adjudicate CodeQL alerts with flow context \citep{iranmanesh2025zerofalse} and adjudicate alerts with multi-run consistency checks \citep{klieber2026adjudicate}. IRIS infers taint specifications for CodeQL and filters its alerts \citep{li2025iris}, LLift triages use-before-initialisation candidates in the Linux kernel \citep{li2024llift}, and LLMDFA synthesises dataflow analyses \citep{wang2024llmdfa}. Our design is closest to IRIS, LLift and RoboDuck: cheap analysis proposes candidates, and a model ranks them. Here, shape localizers and incomplete-fix miners take the place of a full SAST engine, and a single-pass decision model takes the place of a generative reasoner. Overconfidence in LLM security judgements is widespread, and confidence is best used as a relative triage signal \citep{siddiq2026security}, which is how we use it.

\textbf{Small and efficient models for vulnerability detection.} Researchers have specialised small generative models for detection with reasoning and agent scaffolds, including the 7B VulnLLM-R \citep{nie2025vulnllmr} and R2Vul's distilled structured reasoning \citep{weyssow2025r2vul}. Others have compressed models via linear probes \citep{ibanezlissen2025lpass}. Small fine-tuned models do well on synthetic data \citep{bappy2025slmcwe} and can be distilled for C/C++ \citep{wei2025finesec}; for an 8B Llama-class model, fine-tuning matters more than prompting \citep{ouchebara2025llamavuln}. Several results caution that fine-tuning moves thresholds more than understanding: on Linux-kernel C, fine-tuning shifted the output threshold without changing the decision policy \citep{zibaeirad2026calibration}; fine-tuned detectors show high false-positive rates on vulnerable/patched pairs \citep{huang2026semantictrap}; and a classifier on classic code metrics matches state-of-the-art LLMs \citep{weissberg2025metrics}. A leakage-controlled re-evaluation of pre-trained models updates the CodeBERT line \citep{li2025revisiting, feng2020codebert, fu2022linevul}. Vul-RAG adds retrieved vulnerability knowledge \citep{du2024vulrag, kaniewski2026revisitingvulrag}, and LLM4Vuln separates a model's reasoning from its context \citep{sun2024llm4vuln}. Our results support that cautious reading: we use the model to rank, not to decide.

\textbf{Benchmarks and the vulnerable-versus-patched problem.} Risse and Böhme report the closest result to our patched-version AUC: top function-level models cannot tell vulnerable functions from their patched versions \citep{risse2024uncovering}. Paired evaluation has since become standard: PairVul \citep{du2024vulrag}, JitVul \citep{yildiz2025jitvul} and VulnLLMEval \citep{zibaeirad2024vulnllmeval} all report that models struggle to separate a function from its fix. Function-level labels are often undecidable without calling context \citep{risse2025topscore}, precision collapses at realistic base rates \citep{chakraborty2024realvul}, LLM vulnerability analysis leans on surface patterns \citep{li2025svtrusteval}, and fix-derived datasets carry substantial label noise \citep{croft2023dataquality, li2024cleanvul, chen2023diversevul}. PrimeVul \citep{ding2025primevul}, ReVeal \citep{chakraborty2020reveal}, SecVulEval \citep{ahmed2025secvuleval}, CASTLE \citep{dubniczky2025castle} and broad LLM evaluations \citep{khare2025understanding, ullah2024llmscannot, steenhoek2024toerr} likewise paint a pessimistic picture of detection alone. ARVO provides reproducible C/C++ vulnerable/patched builds \citep{mei2026arvo}.

\textbf{Incomplete fixes and recurring vulnerabilities.} Our incomplete-fix miners build on earlier recurring-vulnerability work: vulnerable clone detection \citep{kim2017vuddy}, patch-enhanced vulnerability signatures \citep{xiao2020mvp}, and multi-stage filtering with taint analysis \citep{feng2024fire}; an independent benchmark finds the best of these reach F1 0.55 on harder clones \citep{cao2025recurring}. Patch studies show why incomplete fixes matter: about 6\% of publicly reported security patches needed supplementary patches \citep{woo2025patcheffectiveness}, incomplete fixes in the Linux kernel share recognisable root causes \citep{liu2025incompletelinux}, and many vulnerability fixes take several patches \citep{qi2026multipatch}. Big Sleep framed LLM bug-finding as variant analysis of previously fixed bugs \citep{bigsleep2024sqlite}.

\textbf{Agentic discovery and LLM-assisted fuzzing.} Project Naptime and Big Sleep established tool-equipped agents that verify their own findings \citep{glazunov2024naptime, bigsleep2024sqlite, walker2025summersecurity}; CodeMender and Aardvark extend this to patching \citep{popa2025codemender, openai2025aardvark, venturebeat2025aardvark}. Such agents have also reached the scale of a whole operating-system kernel: OpenAI's run that produced 948 Linux privilege-escalation bugs was priced at about US\$1.2 million in API credits \citep{openai2026paths}. The DARPA AI Cyber Challenge produced complete cyber-reasoning systems combining LLMs with fuzzing and static analysis \citep{darpa2025aixccresults, kim2025atlantis, trailofbits2025buttercup, sheng2025fuzzingbrain, zhang2026sokaixcc}, since made locally deployable \citep{chin2026osscrs} and extended to multi-agent discovery on OSS-Fuzz targets \citep{sheng2026fuzzingbrainv2}. Searching over agent harness designs found browser zero-days with an open-weight model \citep{liu2026agentflow}. Related work includes LLM-written fuzz drivers \citep{lyu2024promptfuzz, zhang2024howeffective, ossfuzzgen, liu2023aifuzzing, google2024levelingup} and agent benchmarks for reproduction and proofs of concept \citep{wang2025cybergym, lee2025secbench, sheng2026fbbench, anthropic2025cyberdefenders}. These systems use generative models throughout; ours confines the model to one ranking pass.

\textbf{Small models and ``system over model'' (concurrent, 2026).} Within days of the Mythos announcement, Fort showed that small and open-weight models, given the isolated vulnerable function, often recognise the same bugs, and argued that the advantage lies in the system rather than the model \citep{fort2026jaggedfrontier, fort2026systemovermodel}. Independent reproductions found the same. Function-level slicing outperformed whole-file prompting for open and frontier models alike, and at function level an open model matched a frontier model \citep{boberg2026needles}. Local open-weight models on one consumer GPU found the target bug but drowned it in false positives until a reachability stage was added \citep{clearbluejar2026systemovermodel}. Other studies reproduced or bounded these results \citep{moczadlo2026vidocmythos, david2026mythosrediscovery, luo2026codeaugur}. Earlier, a frontier model called without scaffolding found a kernel zero-day at a signal-to-noise ratio of about 1:50 \citep{heelan2025o3}; AI analyzers have since been credited with OpenSSL CVEs \citep{aisle2026openssl} and, by the curl maintainer, with a large batch of real fixes \citep{stenberg2025newbreed}. Pesoli et al.~argue that the bottleneck moves from discovery to validation \citep{pesoli2026bugonomics}. Our results support the ``system over model'' argument with a single-pass decision model used mainly for ranking.

\textbf{Memory-safety bug finding.} Memory-safety defects remain the largest class of serious vulnerabilities in C and C++ code \citep{thomas2019msrc70, chromium_memsafety, cisa2023memorysaferoadmaps}. Our oracle is AddressSanitizer \citep{serebryany2012asan, clangasan}, with hardened builds using \texttt{\_FORTIFY\_SOURCE} and stack protectors \citep{glibcfortify, gccinstrumentation}. Coverage-guided fuzzing is the standard way to find these bugs \citep{zalewski_afl, fioraldi2020aflpp, libfuzzer}, and OSS-Fuzz applies it continuously to about a thousand projects \citep{ossfuzz}. Our results suggest that projects outside OSS-Fuzz are where a static approach pays off.

\clearpage
\section{Method: the harness}\label{sec:method}

\subsection{Overview}\label{sec:overview}

By \emph{harness} we mean the whole system around the model. It turns a corpus into a ranked list of candidate functions. Project selection chooses repositories; cheap deterministic signals locate candidate functions within them; a small decision model ranks those candidates rather than accepting or rejecting them, apart from a low reach-score floor on some queues (Section~\ref{sec:decision}). Nothing counts as a finding until it has been reproduced. We call the program that reproduces a suspected defect a \emph{reproducer}.

\subsection{Scope}\label{sec:scope}

The harness targets memory-safety defects in open-source C and C++ software that can be triggered through untrusted input, such as files, network messages or documents. The harness reads source code and, for fix mining, Git history. Its focus is memory accesses, buffer sizes and length calculations. Business logic, authorisation and other classes that need different reasoning fall outside its scope. The search stages---corpus selection, miners, localizers and decision model---use no frontier model. Stronger coding agents take part only in verification; we report its outcomes but do not evaluate that stage. The resource-efficiency claims therefore concern search and prioritisation, not the whole process. Function-level scoring limits what the model sees. A caller, set-up code or build configuration may constrain a candidate that looks unsafe locally; only later investigation can establish those constraints.

\subsection{Where to look: project selection}\label{sec:corpus}

Project selection aims at software whose security affects many downstream users. We tried three strategies.

\begin{itemize}
\tightlist
\item
  \textbf{Curated domains} (file-format, codec, document and network parsers) gave low and mostly minor yield after the first rounds.
\item
  \textbf{GitHub stars} pulled in dead and hobby projects, because stars measure lifetime popularity rather than deployment.
\item
  \textbf{Reach ranking} replaced stars with a reach score over 4,245 C/C++ repositories, built from Repology's cross-distribution packaging data \citep{repology}, Debian reverse dependencies and popularity-contest counts, vcpkg/Conan/Homebrew presence and deps.dev, with a liveness filter.
\end{itemize}

Known fuzzing coverage provides a second criterion. We use OSS-Fuzz membership as an observable indicator \citep{ossfuzz} and treat absence from it as a selection signal, not as proof that a project was never fuzzed. The reach-ranked head is heavily fuzzed and well guarded, so the final campaign rounds drew a \emph{fresh} corpus of widely packaged projects that are mostly outside OSS-Fuzz (Section~\ref{sec:datasets}).

\subsection{Where to look: miners and localizers}\label{sec:localizers}

\textbf{Incomplete-fix miners} read the full Git history. They find commits that add a protective check around an unsafe operation, then look for sibling sites without the check or for checks that fail to exclude the fatal value. A sibling found this way is a hypothesis, not a verdict: its inputs may be constrained elsewhere.

\textbf{Shape localizers} do not depend on fix history. Each is a cheap syntactic detector for one defect shape, such as a fixed-size buffer written with an input-derived length, arithmetic on a length that can wrap before a copy, or an untrusted size reaching a memory operation. A further deterministic test keeps only functions that perform an unsafe memory operation; unlike a model-based cut, this is an explicit part of the candidate criteria. This lets the localizers propose candidates even where no relevant fix exists.

\subsection{The decision: two questions, used to rank}\label{sec:decision}

Each candidate function is scored by Intern-Decision-4B \citep{internlm2026interndecision} with two typed questions: whether the code processes data from an untrusted source (the \emph{reach} question), and a second question matched to its defect shape. The model reads the function's code and returns scores that order the list. Until the localizer pilots of 2026-10-07, every queue also dropped candidates with a reach score below 0.27 (Table~\ref{tbl:ranking}). From the overnight run of 2026-10-07 onward, the fix-miner and fixed-buffer queues kept every candidate and only flagged low reach scores; the other localizer queues kept the 0.27 reach floor through the last scan round on 2026-10-09. No queue used a cutoff on the shape question after the pilots. A candidate kept in the queue is not necessarily read. The review budget determines how far down the list reviewers go; ranking determines what they read first.

\subsection{From lead to finding}\label{sec:proof}

A ranked lead becomes a finding only after verification. Triage and proof used separate, stronger models and human review of the surrounding program context. The decision model did not perform those stages. Every finding had to reproduce as an AddressSanitizer crash in a reproducer and again on a hardened build, and a person signs off before anything is disclosed. A reproduced crash shows failure under the tested conditions. Its security relevance also depends on the real program, including whether untrusted input can reach the failing path. This paper does not evaluate the effectiveness or cost of this stage.

\section{Experimental setup}\label{sec:eval}

\subsection{Research questions}\label{sec:rqs}

\begin{itemize}
\tightlist
\item
  \textbf{RQ1: Classification.} How well do small decision models and other small classifiers separate vulnerable functions from unrelated benign functions, and from their own patched versions? (Section~\ref{sec:model})
\item
  \textbf{RQ2: Prioritisation.} How much do the harness's deterministic signals and the model's ranking concentrate real findings at the top of a limited review budget, compared with random order? (Section~\ref{sec:harness})
\item
  \textbf{RQ3: Recall.} Of known vulnerabilities in projects without any of our findings, how many does the harness place within a repository's review budget, and where are the rest lost? (Section~\ref{sec:recall})
\item
  \textbf{RQ4: Throughput and cost.} How fast is the decision step on data-centre-free hardware, from a desk-side GB10 to a consumer gaming GPU, and how does the harness compare in cost and output with a rule engine? (Section~\ref{sec:speed})
\item
  \textbf{RQ5: Campaign outcomes.} What findings, severities and review outcomes did the campaign produce? (Section~\ref{sec:findings})
\end{itemize}

\subsection{Datasets and study populations}\label{sec:datasets}

\textbf{CVE function set.} 91 vulnerable/patched function pairs from real CVE fixes, plus 246 benign functions from 22 repositories; 71 pairs and 200 benign functions are native C/C++. The vulnerable-versus-benign comparison measures separation from unrelated code; the vulnerable-versus-patched comparison measures separation from the repaired version of the same function. The campaign called this its calibration set, but AUC on it does not by itself establish calibration.

\textbf{Method study.} 339 functions from 82 repositories in the first search rounds, labelled by hand: 15 bug-site functions belonging to 8 confirmed bugs (one bug can span several functions), 301 hand-rejected false positives, and 23 functions with ambiguous labels that were recorded but not scored. We count a bug as kept if any of its sites survives; AUC compares the 8 bugs with the 301 false positives.

\textbf{Ranker evaluation at deployment prevalence.} 3,199 labelled C/C++ records (252 true positives, 2,529 false positives and 418 uncertain) plus a background of 3,000 unread candidates, reweighted to one real bug per 200 candidates; precision@25 is the reviewer budget. The incumbent model chose what was read, biasing the labels toward it.

\textbf{Campaign queues.} The 51 real ranked queues that contain at least one confirmed finding: 44,121 rows, of which 91 map to 78 distinct confirmed findings by repository and function.

\textbf{Known-CVE recall set.} 75 public memory-safety CVEs drawn at random (fixed seed) from NVD by a rule fixed in the selection script before its single run (Appendix~\ref{sec:cvelist}). None is in a project with one of our findings: projects were excluded by name match against the campaign's disclosure status sheet. Most of the 75 projects (66, 54 as the same GitHub repository) appear in the campaign's scan lists, but the campaign scanned them at their October 2026 heads, after these fixes; the recall run checks out each repository at the parent of its fix.

\textbf{Campaign corpus.} The reach-ranked corpus (built 2026-10-07) has 4,245 C/C++ repositories, 311 (7.3\%) of them on OSS-Fuzz. The fresh corpus (2026-10-09) starts from 155,827 Repology projects packaged in at least two families and keeps 2,832 maintained C/C++ repositories within a size limit that are in neither the reach-ranked list nor any earlier scan; 1,178 of them expose an attack surface, and 1,000 were mined. OSS-Fuzz membership was checked through deps.dev for the 1,278 fresh repositories hosted on GitHub at the time: 6 (0.5\%) are enrolled.

\textbf{Campaign findings.} We report the memory-safety findings the campaign confirmed after review, across all corpora it searched (Section~\ref{sec:findings}).

\subsection{Comparisons and procedures}\label{sec:comparisons}

\textbf{Small-classifier comparison.} Six small open models on the native CVE function set, one at a time on one GB10 under vLLM 0.30.1: Granite Guardian 3.1-2B, 3.2-5B and 3.3-8B \citep{ibm2024graniteguardian31, ibm2025graniteguardian32, ibm2025graniteguardian33}, Phi-4-mini-instruct \citep{microsoft2025phi4mini} and xLAM-2 1B and 3B \citep{salesforce2025xlam2}, with Intern-Decision-4B as reference. Only Intern is a trained decision model; the others were read as P(Yes) from a chat prompt or their own guard template. These are configuration baselines rather than a controlled model-family comparison. Each pairs a model with one readout and prompt, sometimes off its native template; none was evaluated as a ranker on real queues. Llama Guard was not run: it is gated, and its largest version exceeds the host's memory headroom.

\textbf{Method study and ranker evaluation.} The method study compares combinations of deterministic tests, models and questions by how many known bugs they keep and how many hand-rejected candidates they remove. We tested thresholds leaving one bug out, but chose the final design with all 8 bugs in view. Its 8-of-8 retention is therefore optimistic; with n = 8, a one-bug difference is noise. The ranker evaluation measures precision@25 at the reweighted deployment prevalence.

\textbf{Ranking ablation.} We re-order each campaign queue by the deployed decision-model score, at random (10,000 seeds), and by a simple heuristic (the localizer's own site count). We count confirmed findings in the top k of each queue, pooled over queues.

\textbf{Recall procedure.} Each known-CVE repository is checked out at the parent of its fix, and the campaign's where-to-look stages run unchanged, including their per-repository caps; every queued row is then scored by the decision model. A CVE counts as found at k if a function the fix changed is in that repository's top k (Appendix~\ref{sec:cvelist}).

\textbf{Rule-engine comparison and hardware portability.} OpenSSL at its master commit of 2026-10-07, a large, heavily fuzzed C repository where the campaign produced no lead. On the same 16-vCPU x86 host we ran Opengrep v1.31.0 with the opengrep-rules C ruleset (16 rules, 3 runs, 16 jobs) and the harness's CPU stages, unchanged. The decision step (278 decisions) was timed on one GB10 under vLLM with batching; on a consumer RTX 4090 in a gaming PC with the same script and engine settings, after a 25-minute CPU and GPU burn-in with checked results; and on a shared laptop's integrated GPU. We measure time, resources and output volume. Neither output was triaged, and CodeQL was not run, so this comparison does not measure detection quality.

\subsection{Metrics and uncertainty}\label{sec:metrics}

\textbf{ROC-AUC} is the probability that a randomly chosen positive receives a higher score than a randomly chosen negative, with ties counted as half; 0.5 is chance and 1.0 perfect separation. We report vulnerable-versus-benign and vulnerable-versus-patched AUC separately. AUC is neither accuracy nor a measure of calibration.

\textbf{Precision@k} is the share of labelled positives among the first \(k\) ranked candidates. It measures the concentration of positives within a fixed review budget. We use budgets of 10, 25 and 40 and state the population and any reweighting with each result.

\textbf{Recall and false-positive rate} at a threshold are TP/(TP + FN) and FP/(FP + TN), over the labelled population evaluated. In the recall study, recall at k is the share of the 75 CVEs whose changed function is in the repository's top k. Neither establishes the recall of the whole campaign, whose missed vulnerabilities are unknown.

\textbf{Review outcomes.} The share of hand-rejected candidates a filter removes measures false-alarm reduction on that set, and the triage keep rate measures how many ranked leads survive triage. Their denominators differ, so they cannot be combined into one pipeline false-positive rate.

\textbf{Throughput} is given in decisions or function chunks per second, with the unit stated for each experiment; rows per second are not comparable across models that answer different numbers of questions with different prompt lengths. Timings exclude investigation and disclosure.

\textbf{Uncertainty.} For Intern-Decision-4B on the CVE function set we report 95\% percentile intervals from a bootstrap that draws each vulnerable/patched pair as one unit and the benign functions separately (1,000 replicates, seed 20261110). The ranking ablation reports 95\% intervals from a bootstrap over queues; the recall study, 95\% intervals on each recall. These intervals describe uncertainty in individual estimates, not in the difference between models or methods.

\subsection{Severity assessment}\label{sec:severity}

Severity is assessed with the Common Vulnerability Scoring System v3.1 \citep{firstcvss31}, whose qualitative bands are none (0), low (0.1--3.9), medium (4.0--6.9), high (7.0--8.9) and critical (9.0--10.0). Each confirmed vulnerability is rated by worst-case impact from its own write-up, with one rubric applied by independent raters, and a blind second rating of 20 findings. Memory-corrupting writes take the code-execution ceiling, except weak writes (a single byte, attacker-uncontrolled bytes, small values inside one structure), which are scored at low confidentiality and integrity impact unless the attacker chooses the offset; out-of-bounds reads and crashes never reach critical. Each rating accounts for reachability from untrusted input. These are campaign scoring assumptions, not general rules implied by a defect class. A reproduced crash demonstrates failure; the assumed impact may go beyond what was demonstrated.

\subsection{Implementation and hardware}\label{sec:system}

\textbf{Decision hosts.} NVIDIA DGX Spark desktop systems (GB10 Grace Blackwell): one GPU, 20 Arm cores and 119.7 GiB of memory shared between CPU and GPU. The campaign used three. NVIDIA's list price for the 128 GB DGX Spark Founders Edition was US\$3,999 at its October 2025 launch and US\$4,699 from February 2026 \citep{technoedge2026dgxspark, nvidia2026dgxsparkprice}; by October 2026 it was listed at US\$6,950 \citep{technoedge2026dgxspark}. We used DGX Spark Founders Edition units, so these list prices describe our hardware. They do not measure campaign cost. Overcommitting unified memory hangs the host rather than failing a request. We therefore loaded models strictly in sequence behind a free-memory guard.

\textbf{Serving.} Throughput experiments used vLLM with batching (a 0.30.1 development build on the GB10; 0.32.0 on the RTX 4090). The 4B ranker's weights are about 8.5 GB in BF16; total runtime memory also depends on input lengths and serving settings.

\textbf{Build host.} Builds and sanitizer runs used an x86-64 build server.

\textbf{Commodity hardware.} We ran the unchanged ranker on a gaming PC's consumer GPU (NVIDIA RTX 4090, 24 GB, under WSL2) and on a laptop's integrated GPU (Apple M4 Max, without vLLM), and compared its scores with the GB10's (Section~\ref{sec:speed}).

\section{Results}\label{sec:results}

\subsection{RQ1: Classification}\label{sec:model}

Alone, a small decision model is a weak detector. On the native C/C++ function set, Intern-Decision-4B separates vulnerable from benign functions with AUC 0.73, but a vulnerable function from its own patch with only 0.56: a fix that changes one token can leave the score almost unchanged (Figure~\ref{fig:pair}). At a fixed threshold of 0.5 under vLLM, it recalls 15\% of vulnerable functions at a 6\% benign false-positive rate. An earlier campaign run used a different serving stack on the same functions and gave 20\% recall and 7\% benign false-positive rate at that threshold, with the same AUCs (0.732 and 0.566). The ranking was stable, but even one model's threshold behaviour moved with how it was served. The campaign's other candidate deciders were within a few points of it (Table~\ref{tbl:deciders}). Against patched code (all 91 pairs), all lie between 0.55 and 0.57; these point estimates do not establish a reliable ordering.

\begin{figure}[t]
\centering
\lstdefinestyle{pairfig}{language=C,basicstyle=\ttfamily\scriptsize,columns=fullflexible,keepspaces=true,upquote=true,breaklines=true,breakindent=1.5em,postbreak=\mbox{\textcolor{gray}{$\hookrightarrow$}\space},escapechar=|,showstringspaces=false,frame=single,rulecolor=\color{gray!50},xleftmargin=2pt,xrightmargin=2pt,aboveskip=2pt,belowskip=0pt,keywordstyle=\bfseries}
\newcommand{\hlvuln}[1]{{\setlength{\fboxsep}{0.5pt}\colorbox{red!25}{\strut #1}}}
\newcommand{\hlfix}[1]{{\setlength{\fboxsep}{0.5pt}\colorbox{green!30}{\strut #1}}}
\begin{minipage}[t]{0.485\linewidth}
\small\textbf{Vulnerable} (parent of fix commit) \hfill decider score \textbf{0.307}
\begin{lstlisting}[style=pairfig]
static cJSON_bool decode_array_index_from_pointer(const unsigned char * const pointer, size_t * const index)
{
    size_t parsed_index = 0;
    size_t position = 0;

    if ((pointer[0] == '0') && ((pointer[1] != '\0') && (pointer[1] != '/')))
    {
        /* leading zeroes are not permitted */
        return 0;
    }

    for (position = 0; (pointer[position] >= '0') && (|\hlvuln{pointer[0]}| <= '9'); position++)
    {
        parsed_index = (10 * parsed_index) + (size_t)(pointer[position] - '0');

    }

    if ((pointer[position] != '\0') && (pointer[position] != '/'))
    {
        return 0;
    }

    *index = parsed_index;

    return 1;
}
\end{lstlisting}
\end{minipage}\hfill
\begin{minipage}[t]{0.485\linewidth}
\small\textbf{Patched} (fix commit 74e1ff4) \hfill decider score \textbf{0.294}
\begin{lstlisting}[style=pairfig]
static cJSON_bool decode_array_index_from_pointer(const unsigned char * const pointer, size_t * const index)
{
    size_t parsed_index = 0;
    size_t position = 0;

    if ((pointer[0] == '0') && ((pointer[1] != '\0') && (pointer[1] != '/')))
    {
        /* leading zeroes are not permitted */
        return 0;
    }

    for (position = 0; (pointer[position] >= '0') && (|\hlfix{pointer[position]}| <= '9'); position++)
    {
        parsed_index = (10 * parsed_index) + (size_t)(pointer[position] - '0');

    }

    if ((pointer[position] != '\0') && (pointer[position] != '/'))
    {
        return 0;
    }

    *index = parsed_index;

    return 1;
}
\end{lstlisting}
\end{minipage}
\caption{One vulnerable/patched pair from the CVE function set: cJSON's \texttt{decode\_\allowbreak array\_\allowbreak index\_\allowbreak from\_\allowbreak pointer} (\texttt{cJSON\_\allowbreak Utils.c}), CVE-2025-57052, before and after the fix. The whole fix is one token in the loop condition (highlighted): the upper-bound digit test reads \texttt{pointer[0]} instead of \texttt{pointer[position]}, so letters after a leading digit are folded into the array index, which the CVE record describes as out-of-bounds access. Intern-Decision-4B, asked the campaign's memory-safety question, scored the two versions 0.307 and 0.294 (the campaign's stored per-function scores). Across the 71 native C/C++ pairs, the vulnerable version scored higher in only 30 (16 ties), and AUC against the patched version was 0.566 in that run (0.564 under vLLM, Table~\ref{tbl:clsperf}). Separating such a pair depends on context outside the function \citep{risse2024uncovering,risse2025topscore}; the score can order a queue, but it cannot decide this function on its own.}
\label{fig:pair}
\end{figure}

\begin{longtable}[]{
  >{\raggedright\arraybackslash}p{(\linewidth - 8\tabcolsep) * \real{0.2900}}
  >{\raggedright\arraybackslash}p{(\linewidth - 8\tabcolsep) * \real{0.2300}}
  >{\raggedright\arraybackslash}p{(\linewidth - 8\tabcolsep) * \real{0.2400}}
  >{\raggedright\arraybackslash}p{(\linewidth - 8\tabcolsep) * \real{0.2400}}}
\caption{Campaign deciders on the CVE function set. A dash marks a measurement not taken. DiffusionGemma-26B-A4B \citep{google2026diffusiongemma} was run as NVIDIA's NVFP4 build \citep{nvidia2026diffusiongemmanvfp4}. The patched column covers all 91 pairs, including 20 in managed languages; on the 71 native pairs Intern scores 0.564 (Table~\ref{tbl:clsperf}).}\label{tbl:deciders}\tabularnewline
\toprule\noalign{}\rowcolor{PaperHeader}
\begin{minipage}[c]{\linewidth}\raggedright\bfseries\fontsize{9}{11}\selectfont
Configuration
\end{minipage} & \begin{minipage}[c]{\linewidth}\raggedright\bfseries\fontsize{9}{11}\selectfont
AUC vs benign (native C/C++)
\end{minipage} & \begin{minipage}[c]{\linewidth}\raggedright\bfseries\fontsize{9}{11}\selectfont
AUC vs patched (all 91 pairs)
\end{minipage} & \begin{minipage}[c]{\linewidth}\raggedright\bfseries\fontsize{9}{11}\selectfont
Throughput on one GB10
\end{minipage} \\
\midrule\noalign{}
\endfirsthead
\toprule\noalign{}\rowcolor{PaperHeader}
\begin{minipage}[c]{\linewidth}\raggedright\bfseries\fontsize{9}{11}\selectfont
Configuration
\end{minipage} & \begin{minipage}[c]{\linewidth}\raggedright\bfseries\fontsize{9}{11}\selectfont
AUC vs benign (native C/C++)
\end{minipage} & \begin{minipage}[c]{\linewidth}\raggedright\bfseries\fontsize{9}{11}\selectfont
AUC vs patched (all 91 pairs)
\end{minipage} & \begin{minipage}[c]{\linewidth}\raggedright\bfseries\fontsize{9}{11}\selectfont
Throughput on one GB10
\end{minipage} \\
\midrule\noalign{}
\endhead
\bottomrule\noalign{}
\endlastfoot
4B + 27B two-tier (rank mean) & 0.76 & 0.57 & --- \\
\rowcolor{PaperStripe}clef-27B & 0.74 & 0.56 & 0.45 chunks/s \\
Intern-Decision-4B $\cdot$ memory-safety question & 0.73 & 0.55 & 4.7 chunks/s \\
\rowcolor{PaperStripe}DiffusionGemma-26B-A4B & 0.69 & 0.55 & --- \\
\end{longtable}

\textbf{Six configuration baselines, measured.} The comparison concerns scoring configurations rather than model families (Section~\ref{sec:comparisons}). None separates a vulnerable function from its patch better (Table~\ref{tbl:clsperf}): AUC against the patched version ranged from 0.50 to 0.62. Against benign code, xLAM-2-1B (0.77) and Intern (0.73) scored highest. Intern's own 95\% interval is 0.66--0.79 against benign code and 0.52--0.61 against patched code (pair-preserving bootstrap, Section~\ref{sec:metrics}). xLAM-2-1B's 0.77 lies within the first interval. This describes the uncertainty in Intern's estimate on 71 vulnerable functions; it is not a test of the difference between the two models. The interval comes from the campaign's scores for the same model and functions (AUC 0.732 and 0.566). Table~\ref{tbl:clsperf} shows the later vLLM run, whose point estimates agree to 0.002. The threshold of 0.5 behaved very differently across models: Granite Guardian 3.1-2B flagged 84\% of benign code, while xLAM-2-3B flagged nothing. Table~\ref{tbl:clsrun} gives sizes, readouts and throughput.

\begin{longtable}[]{
  >{\raggedright\arraybackslash}p{(\linewidth - 10\tabcolsep) * \real{0.2300}}
  >{\raggedright\arraybackslash}p{(\linewidth - 10\tabcolsep) * \real{0.2000}}
  >{\raggedright\arraybackslash}p{(\linewidth - 10\tabcolsep) * \real{0.2000}}
  >{\raggedright\arraybackslash}p{(\linewidth - 10\tabcolsep) * \real{0.1700}}
  >{\raggedright\arraybackslash}p{(\linewidth - 10\tabcolsep) * \real{0.2000}}}
\caption{Small classifiers on the native C/C++ function set (71 vulnerable, 71 patched, 200 benign), one GB10 under vLLM. Intern's 95\% intervals are given in the text.}\label{tbl:clsperf}\tabularnewline
\toprule\noalign{}\rowcolor{PaperHeader}
\begin{minipage}[c]{\linewidth}\raggedright\bfseries\fontsize{9}{11}\selectfont
Model
\end{minipage} & \begin{minipage}[c]{\linewidth}\raggedright\bfseries\fontsize{9}{11}\selectfont
AUC vs benign
\end{minipage} & \begin{minipage}[c]{\linewidth}\raggedright\bfseries\fontsize{9}{11}\selectfont
AUC vs patched
\end{minipage} & \begin{minipage}[c]{\linewidth}\raggedright\bfseries\fontsize{9}{11}\selectfont
Recall at 0.5
\end{minipage} & \begin{minipage}[c]{\linewidth}\raggedright\bfseries\fontsize{9}{11}\selectfont
Benign FPR at 0.5
\end{minipage} \\
\midrule\noalign{}
\endfirsthead
\toprule\noalign{}\rowcolor{PaperHeader}
\begin{minipage}[c]{\linewidth}\raggedright\bfseries\fontsize{9}{11}\selectfont
Model
\end{minipage} & \begin{minipage}[c]{\linewidth}\raggedright\bfseries\fontsize{9}{11}\selectfont
AUC vs benign
\end{minipage} & \begin{minipage}[c]{\linewidth}\raggedright\bfseries\fontsize{9}{11}\selectfont
AUC vs patched
\end{minipage} & \begin{minipage}[c]{\linewidth}\raggedright\bfseries\fontsize{9}{11}\selectfont
Recall at 0.5
\end{minipage} & \begin{minipage}[c]{\linewidth}\raggedright\bfseries\fontsize{9}{11}\selectfont
Benign FPR at 0.5
\end{minipage} \\
\midrule\noalign{}
\endhead
\bottomrule\noalign{}
\endlastfoot
Intern-Decision-4B & 0.731 & 0.564 & 15\% & 6\% \\
\rowcolor{PaperStripe}xLAM-2-1b-fc-r & 0.773 & 0.590 & 73\% & 32\% \\
xLAM-2-3b-fc-r & 0.688 & 0.544 & 0\% & 0\% \\
\rowcolor{PaperStripe}Phi-4-mini-instruct & 0.666 & 0.617 & 79\% & 50\% \\
Granite Guardian 3.1-2B & 0.672 & 0.555 & 99\% & 84\% \\
\rowcolor{PaperStripe}Granite Guardian 3.2-5B & 0.561 & 0.504 & 34\% & 23\% \\
Granite Guardian 3.3-8B & 0.599 & 0.589 & 17\% & 19\% \\
\end{longtable}

\begin{table}[!htbp]
\centering\small
\caption{Configurations and throughput in the small-classifier comparison (one GB10, vLLM 0.30.1). \mbox{\textsuperscript{*} Intern} answers two fields per row with longer prompts, so its rows/s is not comparable.}\label{tbl:clsrun}
\begin{tabular}{>{\raggedright\arraybackslash}p{0.35\textwidth}>{\raggedright\arraybackslash}p{0.13\textwidth}>{\raggedright\arraybackslash}p{0.275\textwidth}>{\raggedleft\arraybackslash}p{0.12\textwidth}}
\toprule
\rowcolor{PaperHeader}\textbf{Model} & \textbf{Parameters} & \textbf{Readout} & \textbf{Rows/s} \\
\midrule
Intern-Decision-4B & 4.5B & decision readout & 4.7\textsuperscript{*} \\
\rowcolor{PaperStripe}xLAM-2-1b-fc-r & 1.5B & Yes/No & 49.1 \\
xLAM-2-3b-fc-r & 3.1B & Yes/No & 25.3 \\
\rowcolor{PaperStripe}Phi-4-mini-instruct & 3.8B & Yes/No & 23.1 \\
Granite Guardian 3.1-2B & 2.5B & guard template & 18.8 \\
\rowcolor{PaperStripe}Granite Guardian 3.2-5B & 5.8B & guard template & 10.0 \\
Granite Guardian 3.3-8B & 8.2B & guard template & 7.1 \\
\bottomrule
\end{tabular}
\end{table}

\textbf{Why this ranker.} In the method study, the 4B model had the \emph{lowest} best-question AUC of four models (0.653 against 0.728--0.749). Yet its reach question was the only model question that kept all 8 bugs out of sample; a 27B alternative cost about ten times more per chunk and ranked no better inside the filtered pool.

\subsection{RQ2: Prioritisation}\label{sec:harness}

\textbf{Candidate yield in the localizer pilots.} The pilots describe candidate yield by configuration (Table~\ref{tbl:ranking}); they do not isolate the effect of ranking. For the length-wrap localizer, a random sample of 30 candidates held 2 leads (6.7\%). The top 30 after ranking held 9 leads and 1 near-lead (30--33\%). For the untrusted-size localizer, the figures were 3.3\% and 24--36\% in the top 25. For the fixed-buffer localizer, a random sample of 40 from its first version held 1 lead (2.5\%); the top 40 of its revised version held 14 leads and 3 near-leads (35--42.5\%). That list includes the repository used to validate the localizer. Without it, 4 of 24 (16.7\%) were leads. These shares measure the queue at triage, not final findings. A random sample is not a random ordering at the same budget. The columns also differ in localizer version, deterministic test and reach cut, so we reserve \emph{ranking lift} for the within-queue comparisons below.

\begin{longtable}[]{
  >{\raggedright\arraybackslash}p{(\linewidth - 10\tabcolsep) * \real{0.2200}}
  >{\raggedright\arraybackslash}p{(\linewidth - 10\tabcolsep) * \real{0.2000}}
  >{\raggedright\arraybackslash}p{(\linewidth - 10\tabcolsep) * \real{0.1300}}
  >{\raggedright\arraybackslash}p{(\linewidth - 10\tabcolsep) * \real{0.1900}}
  >{\raggedright\arraybackslash}p{(\linewidth - 10\tabcolsep) * \real{0.2600}}}
\caption{Candidate yield in the localizer pilots, by configuration, at triage (descriptive). Raw rates are hit rates in a random sample of the localizer's candidates; ranked rates are hit rates in the top k after the deterministic test, a reach cut at 0.27 and ranking by the shape question. Ranges count leads only and leads plus near-leads; they are not confidence intervals.}\label{tbl:ranking}\tabularnewline
\toprule\noalign{}\rowcolor{PaperHeader}
\begin{minipage}[c]{\linewidth}\raggedright\bfseries\fontsize{9}{11}\selectfont
Candidate source
\end{minipage} & \begin{minipage}[c]{\linewidth}\raggedright\bfseries\fontsize{9}{11}\selectfont
Random sample
\end{minipage} & \begin{minipage}[c]{\linewidth}\raggedright\bfseries\fontsize{9}{11}\selectfont
Raw rate
\end{minipage} & \begin{minipage}[c]{\linewidth}\raggedright\bfseries\fontsize{9}{11}\selectfont
Top k after ranking
\end{minipage} & \begin{minipage}[c]{\linewidth}\raggedright\bfseries\fontsize{9}{11}\selectfont
Ranked rate
\end{minipage} \\
\midrule\noalign{}
\endfirsthead
\toprule\noalign{}\rowcolor{PaperHeader}
\begin{minipage}[c]{\linewidth}\raggedright\bfseries\fontsize{9}{11}\selectfont
Candidate source
\end{minipage} & \begin{minipage}[c]{\linewidth}\raggedright\bfseries\fontsize{9}{11}\selectfont
Random sample
\end{minipage} & \begin{minipage}[c]{\linewidth}\raggedright\bfseries\fontsize{9}{11}\selectfont
Raw rate
\end{minipage} & \begin{minipage}[c]{\linewidth}\raggedright\bfseries\fontsize{9}{11}\selectfont
Top k after ranking
\end{minipage} & \begin{minipage}[c]{\linewidth}\raggedright\bfseries\fontsize{9}{11}\selectfont
Ranked rate
\end{minipage} \\
\midrule\noalign{}
\endhead
\bottomrule\noalign{}
\endlastfoot
Fixed-buffer localizer & 40 (first version) & 2.5\% & top 40 (revised version) & 35--42.5\% (16.7\% without the validation repository) \\
\rowcolor{PaperStripe}Length-wrap localizer & 30 & 6.7\% & top 30 & 30--33\% \\
Untrusted-size localizer & 30 & 3.3\% & top 25 & 24--36\% \\
\end{longtable}

\textbf{Ranking lift against random order.} The recall study provides a cleaner check because its labels come from public CVE fixes (Section~\ref{sec:recall}). Within the same queues, the model placed the fixed functions in the top 10, 25 and 40 for 9, 16 and 17 CVEs, against 7.0, 11.1 and 13.4 expected at random: a lift of 1.3--1.4 times. On the campaign's own queues the observed lift is larger. Across the 51 campaign queues with a confirmed finding, the model's order put 24 confirmed findings in the top 10, against 3.2 expected at random and 6 for the site-count heuristic. That is a 7.5-fold lift (95\% CI 4.3--14.1). At the top 25 the counts were 32, 6.7 and 13 (4.8-fold, CI 2.9--8.0); at the top 40, 37, 9.5 and 23. Confirmed findings sat at a mean percentile rank of 0.14 under the model, 0.37 under the heuristic and 0.5 at random. The observed lift is a selection-biased upper bound. The 51 queues were chosen because they contain a confirmed finding, reviewers read from the top of these same rankings, and rejected rows rank even higher (0.12).

\textbf{One question after one free test.} In the method study every model reached AUC 0.65--0.75, and cascades of model judgements lost bugs (the best two-stage cascade kept 3 of 8). The best design placed a cheap model after a free deterministic unsafe-operation test and asked one reach question. It cut the hand-rejected false alarms from 301 to 132 ($-$56\%) and kept all 8 bugs.

\textbf{Ranking, not thresholding.} The same model failed repeatedly as a gate. A cutoff at 0.5 kept none of three raw fixed-buffer hits; a rounded published threshold silently dropped 2 of 8 known bugs; and inside a band already above 0.7 the score carried almost no information (mean 0.7782 for kept leads, 0.7684 for discarded ones).

\textbf{Where to look matters most.} After triage, the under-fuzzed corpus kept 20.1\% of ranked leads with reach $\geq$0.7, against 4--12\% for the reach-ranked rounds. The incomplete-fix class (a guard added in one place and missed in its twin) was the most reliable source of real findings. Mature, heavily fuzzed code (cryptography, DNS, fonts, the most depended-on libraries) yielded nothing in our campaign. This describes our chosen targets, not those categories in general.

\textbf{At the deployment base rate.} At a reweighted prevalence of one real bug per 200 candidates, the 4B model reached precision@25 of 0.092, versus 0.062 for a 27B alternative. Fusing the two scored 0.26, but the gain was not proven at a 25--50-read budget, so we did not adopt it.

\subsection{RQ3: Recall on known CVEs}\label{sec:recall}

On 75 public CVEs from projects without any of our findings (Table~\ref{tbl:recall}), the fixed function reached its repository's top 10 for 9 (12\%), the top 25 for 16 (21\%) and the top 40 for 17 (23\%). Almost all of the loss happens before ranking. The changed function appeared in the chunked corpus for 72 of 75 CVEs. A localizer or miner flagged it for only 29, and 18 survived the per-repository caps to be scored. Once scored, the model placed 16 of 18 in the top 25. Random order would place about 11 in the top 25 of these short queues (a median of 60 functions per repository), so ranking adds less here than on the campaign's long queues. The localizers target specific write shapes. They flagged 18 of 41 out-of-bounds writes, but only 7 of 19 out-of-bounds reads and 2 of 8 use-after-free defects.

\begin{longtable}[]{
  >{\raggedright\arraybackslash}p{(\linewidth - 6\tabcolsep) * \real{0.4500}}
  >{\raggedright\arraybackslash}p{(\linewidth - 6\tabcolsep) * \real{0.2700}}
  >{\raggedright\arraybackslash}p{(\linewidth - 6\tabcolsep) * \real{0.2800}}}
\caption{Recall on 75 known CVEs. 95\% intervals for recall at 10, 25 and 40: 5--20\%, 12--31\% and 13--32\%.}\label{tbl:recall}\tabularnewline
\toprule\noalign{}\rowcolor{PaperHeader}
\begin{minipage}[c]{\linewidth}\raggedright\bfseries\fontsize{9}{11}\selectfont
Stage (n = 75 CVEs)
\end{minipage} & \begin{minipage}[c]{\linewidth}\raggedright\bfseries\fontsize{9}{11}\selectfont
Count
\end{minipage} & \begin{minipage}[c]{\linewidth}\raggedright\bfseries\fontsize{9}{11}\selectfont
Share
\end{minipage} \\
\midrule\noalign{}
\endfirsthead
\toprule\noalign{}\rowcolor{PaperHeader}
\begin{minipage}[c]{\linewidth}\raggedright\bfseries\fontsize{9}{11}\selectfont
Stage (n = 75 CVEs)
\end{minipage} & \begin{minipage}[c]{\linewidth}\raggedright\bfseries\fontsize{9}{11}\selectfont
Count
\end{minipage} & \begin{minipage}[c]{\linewidth}\raggedright\bfseries\fontsize{9}{11}\selectfont
Share
\end{minipage} \\
\midrule\noalign{}
\endhead
\bottomrule\noalign{}
\endlastfoot
Changed function in the chunked corpus & 72 & 96\% \\
\rowcolor{PaperStripe}Flagged by a localizer or miner (before caps) & 29 & 39\% \\
Scored (survived the per-repository caps) & 18 & 24\% \\
\rowcolor{PaperStripe}In the repository's top 10 / 25 / 40 & 9 / 16 / 17 & 12\% / 21\% / 23\% \\
Expected at random, same queues (top 10 / 25 / 40) & 7.0 / 11.1 / 13.4 & \\
\end{longtable}

\subsection{RQ4: Throughput and cost}\label{sec:speed}

\textbf{A desk-side GB10.} With vLLM batching on one GB10, Intern-Decision-4B made 278 decisions in 63 s (4.4 decisions per second, 8.0 GiB of weights). During the campaign it scored 3.5--5.5 function chunks per second at 95\% GPU utilisation, against 0.45--0.89 for a dense 27B decider.

\textbf{A consumer gaming PC.} On an RTX 4090 in a gaming PC (24 GB, under WSL2), the same script made the same 278 OpenSSL decisions in 25--29 s: 9.8 decisions per second on the slowest of three passes, 2.2 times the GB10, after a 97 s cold start. The scores matched the GB10's closely: 275 of 278 were within 0.02, all 278 were on the same side of 0.5, and 277 of 278 CWE choices were identical. The run used vLLM 0.32.0. Device memory peaked at 15.2 GiB, including the desktop's own use; a 25-minute burn-in beforehand produced no mismatches or hardware errors. The decision step thus ran faster on the gaming PC than on the GB10.

\textbf{Even a laptop.} On a shared laptop's integrated GPU without vLLM, the same decision step took about 10 minutes; 270 of 278 scores were within 0.02 of the GB10's. This is a floor, not the harness's speed.

\textbf{Scale.} All 4,245 reach-ranked repositories were scanned by both the fix miners and the shape localizers by 2026-10-09. Mining ran at 900--1,100 repositories per hour across two systems; one 300-repository batch took 18 minutes to mine and 4 minutes to score. These timings cover the search stages only.

\textbf{Against a rule engine.} On one repository, Opengrep is about ten times faster end to end (Table~\ref{tbl:cost}). Nearly all of the gap comes from the harness's single-process CPU stages, which the campaign parallelised across repositories rather than within one. The GPU step took about one minute on the GB10 and under half a minute on the RTX 4090. Their outputs also differ. One hygiene rule dominated Opengrep's 554 warnings (\texttt{memset} that may be optimised away: 468), while the harness produced a ranked list of 139 rows. They also looked mostly at different code: only 33 of Opengrep's warnings fall inside the 121 functions the harness sent to the model.

\begin{table}[!htbp]
\centering\small
\caption{One-repository cost comparison on OpenSSL. CPU stages ran on the same 16-vCPU x86 host. The harness's 139 ranked rows cover 121 distinct functions. Neither output was triaged, so this is not a quality comparison.}\label{tbl:cost}
\begin{tabular}{>{\raggedright\arraybackslash}p{(\linewidth - 6\tabcolsep) * \real{0.3800}}>{\raggedright\arraybackslash}p{(\linewidth - 6\tabcolsep) * \real{0.2800}}>{\raggedright\arraybackslash}p{(\linewidth - 6\tabcolsep) * \real{0.3400}}}
\toprule
\rowcolor{PaperHeader} & \textbf{Opengrep v1.31.0, C rules} & \textbf{Harness + 4B decision model} \\
\midrule
CPU wall time (same 16-vCPU host) & 27--31 s at 16 jobs & $\approx$268 s, single process \\
\rowcolor{PaperStripe}Decision step, one GB10, vLLM & --- & 278 decisions in 63 s (4.4/s) \\
Decision step, consumer RTX 4090, vLLM & --- & 278 decisions in 25--29 s (9.8/s) \\
\rowcolor{PaperStripe}Decision step, laptop integrated GPU (floor) & --- & 278 decisions in $\approx$10 min \\
End to end & $\approx$0.5 min & $\approx$5.5 min (GB10); $\approx$5 min (RTX 4090) \\
\rowcolor{PaperStripe}Output & 554 warnings in 227 files & 139 ranked rows; 5 where both questions agree \\
\bottomrule
\end{tabular}
\end{table}

\textbf{Context: the cost of frontier-scale search.} For context, OpenAI reported a single run of a cyber-specialised frontier model against the Linux kernel that produced 948 local privilege-escalation bugs. It gives the cost as US\$1,213,430 in API credits, or about US\$1,280 per bug, and describes that figure as an upper bound \citep{openai2026paths}. Both efforts look for exploitable defects in C systems code, but they spend compute differently. OpenAI's figure covers a frontier model doing search, validation and exploit development. In our harness, search and ranking run on three desk-side systems with a list price of US\$4,000--7,000 each, or on a gaming PC; only the read candidates go to stronger models. The efforts are not comparable per finding. Their targets differ (one kernel versus thousands of user-space packages), as does the bar for a finding (a working root exploit versus a reproduced memory-safety crash); we also do not measure the cost of our verification stage. We have no end-to-end cost accounting. Total model-agent spend, verification cost and reviewer time were not recorded, and our GPUs are owned hardware with no metered cost. Whether a cheap search stage lowers the total cost of finding a vulnerability is therefore a hypothesis this paper motivates but does not test.

\subsection{RQ5: Campaign outcomes}\label{sec:findings}

\begin{figure}[t]
\centering
\definecolor{sevcrit}{HTML}{B00000}
\definecolor{sevhigh}{HTML}{D9730D}
\definecolor{sevmed}{HTML}{E6B800}
\definecolor{sevlow}{HTML}{CDE8C5}
\begin{tikzpicture}
\begin{axis}[
  xbar, bar width=11pt, bar shift=0pt, width=0.62\linewidth, height=4.2cm,
  xmin=0, xmax=175, enlarge y limits=0.18,
  symbolic y coords={Low,Medium,High,Critical},
  ytick={Low,Medium,High,Critical},
  yticklabels={Low (0.1--3.9),Medium (4.0--6.9),High (7.0--8.9),Critical (9.0--10.0)}, y tick label style={font=\small},
  xlabel={Confirmed findings}, xlabel style={font=\small}, x tick label style={font=\small},
  axis x line*=bottom, axis y line*=left, xmajorgrids, grid style={gray!25},
  nodes near coords, nodes near coords style={font=\small, anchor=west},
  every node near coord/.append style={xshift=1pt},
]
\addplot[draw=black, fill=sevcrit] coordinates {(12,Critical)};
\addplot[draw=black, fill=sevhigh] coordinates {(148,High)};
\addplot[draw=black, fill=sevmed] coordinates {(23,Medium)};
\addplot[draw=black, fill=sevlow] coordinates {(4,Low)};
\end{axis}
\end{tikzpicture}
\caption{Worst-case CVSS v3.1 ratings of the 187 confirmed findings, by severity class (score range in parentheses), from Critical to Low. Each bar is labelled with its count; the four classes total 187 findings. 160 of the 187 (86\%) are rated critical or high under the worst-case rubric of Section~\ref{sec:severity}.}
\label{fig:severity}
\end{figure}
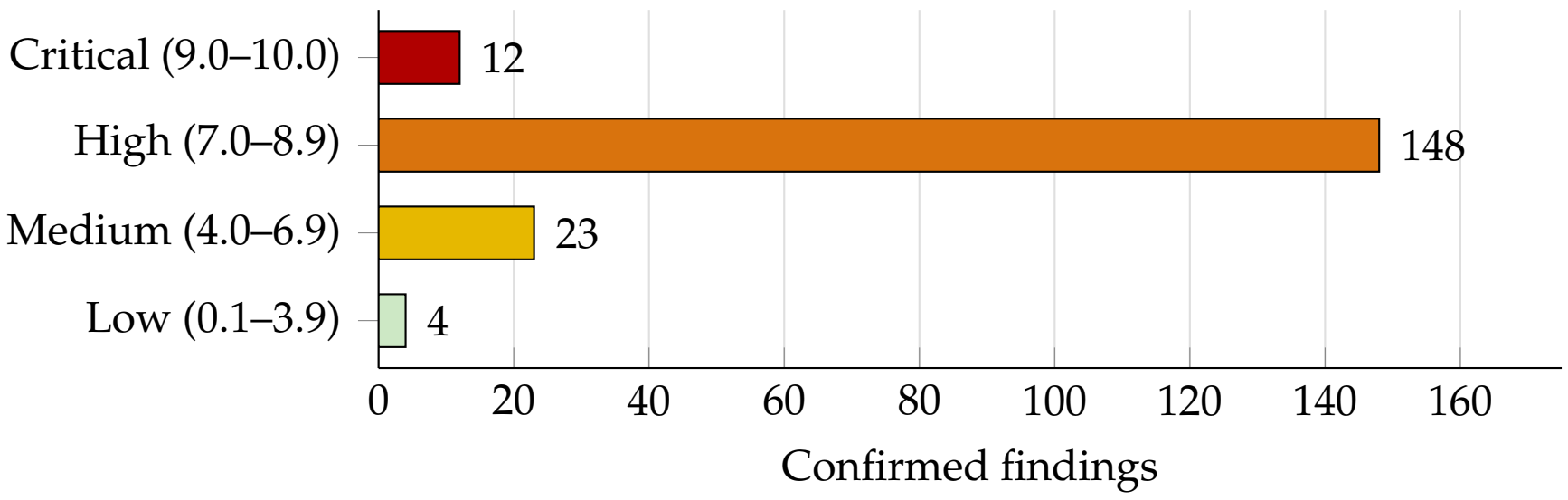

\textbf{Findings and severity.} The campaign confirmed 187 memory-safety findings. Each reproduced as an AddressSanitizer crash in a reproducer and again on a hardened build, and each was accepted in review (Section~\ref{sec:proof}); reproduction in a reproducer does not by itself show reachability through a project's shipped entry point. Figure~\ref{fig:severity} gives their worst-case CVSS v3.1 ratings. Under the rubric of Section~\ref{sec:severity}, 160 of the 187 (86\%) are rated critical or high. All 12 critical ratings are memory-corrupting writes reachable without authentication from the network or from clients that run unattended. The blind second rating agreed on all 20 findings it covered. These ratings assume worst-case impact; they do not establish demonstrated code execution for every finding.

\textbf{False positives.} The record offers several measures, each with a different denominator. They cannot be combined into one false-positive rate.

\begin{itemize}
\tightlist
\item
  \textbf{Candidate yield (triage queue, localizer pilots; descriptive):} 6.7\% $\rightarrow$ 30--33\% in the top 30 (length wrap), 3.3\% $\rightarrow$ 24--36\% in the top 25 (untrusted size), and 2.5\% $\rightarrow$ 35--42.5\% in the top 40 (fixed buffer; 16.7\% without its validation repository).
\item
  \textbf{Deterministic test plus one question (method study):} hand-rejected false alarms 301 $\rightarrow$ 132 ($-$56\%), all 8 known bugs kept.
\item
  \textbf{Classification (known CVEs):} at a fixed 0.5, recall 15\% at a 6\% benign false-positive rate under vLLM (20\% and 7\% in the campaign's own run of the same model), which is why scores are used to rank rather than to decide.
\item
  \textbf{Triage keep rate:} 20.1\% of leads with reach $\geq$0.7 kept in the under-fuzzed corpus, against 4--12\% in reach-ranked rounds.
\end{itemize}

\textbf{What is hard.} Per-function questions miss context outside the function: a caller's check, a build flag that compiles the code out, or a value fixed by set-up code. Applied to smart contracts, the same approach produced no confirmed leads (61 of 61 false positives); real exploits there need cross-function reasoning. Business logic, authorisation and cross-function flows remain outside what a single-function decision model does well.

\section{Responsible disclosure and ethics}\label{sec:disclosure}

This paper names no project, function or defect affected by our findings; the appendix lists only public CVEs, already fixed, used to measure recall. All work ran in contained local builds. Nothing is sent without human sign-off, and the operator declined to report findings in one project of unclear legitimacy.

\textbf{Disclosure.} All confirmed findings are being disclosed to maintainers under coordinated disclosure before publication. We report findings project by project through each project's security channel and work with maintainers on fixes. We will share details of some findings once fixes are available. Until then, we report no per-project or per-fix status.

\textbf{Dual use.} These methods lower the cost of finding bugs. We publish the design, but withhold the corpora, lead lists, harnesses and proofs of concept: maintainers and defenders benefit from knowing that this kind of search is now cheap.

\section{Limitations}\label{sec:limitations}

\begin{itemize}
\tightlist
\item
  \textbf{Triage and proof used stronger models.} The corpus, miners, localizers and decision model use no frontier model. Triage and proof used separate, stronger coding-agent models; we did not measure their cost. A fully small-model pipeline is untested.
\item
  \textbf{Recall is bounded by where to look.} On known CVEs, the harness found about one in five within a repository's top 25. Most misses were never flagged by a localizer. The localizers cover specific write shapes. Reads, use-after-free and logic errors need further localizers; larger per-repository caps would recover some flagged functions (11 of 75 in the recall study).
\item
  \textbf{The ranking ablation is an upper bound.} Reviewers read from the model's own ranking, making low-ranked findings less likely to be confirmed at all.
\item
  \textbf{Small samples and biased labels.} The method study has 8 bugs, and the ranker evaluation's labels were chosen by the incumbent model.
\item
  \textbf{Small-classifier comparison.} Only Intern is a trained decision model, and the other six models were adapted. The sample contains 71 vulnerable functions; ranking on real queues was measured for Intern only.
\item
  \textbf{Only a cost comparison with rule engines,} on one repository; detection quality against Opengrep or CodeQL is future work.
\item
  \textbf{Scope.} C/C++ memory safety only.
\end{itemize}

\section{Conclusion}\label{sec:conclusion}

For the model and harness we studied, a small, fast decision model helped find vulnerabilities at scale because the harness around it did the searching. On its own, the 4B model barely tells a vulnerable function from its fix. The harness decides where to look with incomplete-fix miners, shape localizers and a corpus chosen for reach and for the absence of fuzzing; it uses the model to rank rather than to decide, apart from a low reach-score floor on some queues. For known CVEs from projects without any of our findings, the model placed fixed functions within the review budget 1.3--1.4 times as often as random order of the same queues. On the campaign's queues it put 7.5 times more confirmed findings in the top 10 than random order, an observed upper bound. Across three desk-side GPU systems, the harness searched more than 4,000 projects, and the campaign confirmed 187 memory-safety findings. The search needed no frontier model, and its decision step runs at 9.8 decisions per second on a consumer gaming GPU. On known CVEs, recall is bounded mainly by where the harness looks. That is where the next gains lie. Our interpretation, which this study does not test as a controlled comparison of model sizes, is that careful harness design, more than model size, is what makes early, large-scale vulnerability finding affordable for defenders and maintainers.

\section{Acknowledgements}\label{sec:ack}

We thank Cyrex for supporting this research and NVIDIA for the GB10 (DGX Spark-class) platform used in the experiments. We thank the open-source maintainers who triaged, discussed and fixed our reports; we do not name them until their fixes are public. We thank the InternLM team for releasing Intern-Decision, the Qwen team for the base models it builds on, and the vLLM project for the open serving stack the experiments ran on. The corpus selection relied on public data from Repology, the Debian popularity contest and the OSS-Fuzz project list.

\begingroup\catcode`\_=12
\bibliographystyle{unsrtnat}
\bibliography{REFERENCES-print}
\endgroup
\clearpage
\appendix
\small
\section{Known-CVE recall set}\label{sec:cvelist}

\textbf{Selection rule.} The selection script fixed the rule and random seed before its single run. The written evaluation plan specified the dates, CWE set, C/C++ fix, two-family packaging, status-sheet exclusion, 1--3-function limit and a seeded draw of 75. The script added one CVE per repository, the 2 GB size limit and the seed value. We used only public sources.

\begin{enumerate}
\def\labelenumi{\arabic{enumi}.}
\tightlist
\item
  NVD CVEs published from 2023-01-01 to 2026-09-30 whose CWE is one of 787, 121, 122, 125, 416, 190, 191, 131 or 119, and which reference a fix commit on GitHub.
\item
  The fix changes C or C++ files and 1--3 functions (counted from the diff's hunk headers).
\item
  The project is packaged in at least two Repology families, and its name does not match a project on the campaign's disclosure status sheet (case- and punctuation-insensitive; names of five or more characters also match as substrings).
\item
  One CVE per repository; repository size at most 2 GB.
\item
  Candidates are shuffled with seed 20261028 and taken in order until 75 are accepted.
\end{enumerate}

Of 519 CVEs with a GitHub fix commit, 290 were considered in seeded order to accept 75. The main reasons for rejection were: a repository already selected, fewer than two Repology families, no C/C++ change, a project on the status sheet, or a fix spanning 0 or more than 3 functions. The accepted set is 41 CWE-787, 19 CWE-125, 8 CWE-416, 5 CWE-190, 1 CWE-119 and 1 CWE-121.

\textbf{Targets and counting.} Each repository is checked out at the parent of the fix commit. Target functions are those whose code at that commit overlaps a line changed by the fix. Windows of one large function count as that function. A CVE is found at k if any target function is among the repository's k highest-scored functions, with ties counted against us. Because hunk headers can name a neighbouring function, we report the line-based targets. Counting by hunk-header names gives 10, 16 and 17 at 10, 25 and 40. Counting only the exact window containing the change gives 8, 11 and 12.

\begin{longtable}[]{
  >{\raggedright\arraybackslash}p{(\linewidth - 8\tabcolsep) * \real{0.2105}}
  >{\raggedright\arraybackslash}p{(\linewidth - 8\tabcolsep) * \real{0.0789}}
  >{\raggedright\arraybackslash}p{(\linewidth - 8\tabcolsep) * \real{0.5000}}
  >{\raggedright\arraybackslash}p{(\linewidth - 8\tabcolsep) * \real{0.2105}}}
\caption{The 75 public CVEs of the recall evaluation, with the fix commit used (the repository is checked out at its parent). Repositories are as referenced by NVD; one reference points to a GitHub mirror of the project.}\label{tbl:cvelist}\tabularnewline
\toprule\noalign{}\rowcolor{PaperHeader}
\begin{minipage}[c]{\linewidth}\raggedright\bfseries\fontsize{9}{11}\selectfont
CVE
\end{minipage} & \begin{minipage}[c]{\linewidth}\raggedright\bfseries\fontsize{9}{11}\selectfont
CWE
\end{minipage} & \begin{minipage}[c]{\linewidth}\raggedright\bfseries\fontsize{9}{11}\selectfont
Repository
\end{minipage} & \begin{minipage}[c]{\linewidth}\raggedright\bfseries\fontsize{9}{11}\selectfont
Fix commit
\end{minipage} \\
\midrule\noalign{}
\endfirsthead
\caption[]{Continued from the previous page}\tabularnewline
\toprule\noalign{}\rowcolor{PaperHeader}
\begin{minipage}[c]{\linewidth}\raggedright\bfseries\fontsize{9}{11}\selectfont
CVE
\end{minipage} & \begin{minipage}[c]{\linewidth}\raggedright\bfseries\fontsize{9}{11}\selectfont
CWE
\end{minipage} & \begin{minipage}[c]{\linewidth}\raggedright\bfseries\fontsize{9}{11}\selectfont
Repository
\end{minipage} & \begin{minipage}[c]{\linewidth}\raggedright\bfseries\fontsize{9}{11}\selectfont
Fix commit
\end{minipage} \\
\midrule\noalign{}
\endhead
\bottomrule\noalign{}
\endlastfoot
CVE-2021-33796 & 416 & ccxvii/mujs & \texttt{7ef066a3bb95} \\
\rowcolor{PaperStripe}CVE-2021-45985 & 787 & lua/lua & \texttt{cf613cdc6fa3} \\
CVE-2022-28071 & 416 & radareorg/radare2 & \texttt{65448811e5b9} \\
\rowcolor{PaperStripe}CVE-2022-48468 & 190 & protobuf-c/protobuf-c & \texttt{ec3d900001a1} \\
CVE-2023-0341 & 787 & editorconfig/editorconfig-core-c & \texttt{41281ea82fbf} \\
\rowcolor{PaperStripe}CVE-2023-2804 & 787 & libjpeg-turbo/libjpeg-turbo & \texttt{9f756bc67a84} \\
CVE-2023-4255 & 787 & tats/w3m & \texttt{edc602651c50} \\
\rowcolor{PaperStripe}CVE-2023-4755 & 416 & gpac/gpac & \texttt{895ac12da168} \\
CVE-2023-5217 & 787 & webmproject/libvpx & \texttt{3fbd1dca6a4d} \\
\rowcolor{PaperStripe}CVE-2023-5341 & 416 & imagemagick/imagemagick & \texttt{aa673b2e4def} \\
CVE-2023-23457 & 119 & upx/upx & \texttt{779b648c5f6a} \\
\rowcolor{PaperStripe}CVE-2023-25567 & 125 & gssapi/gss-ntlmssp & \texttt{025fbb756d44} \\
CVE-2023-29418 & 125 & kspalaiologos/bzip3 & \texttt{aae16d107f80} \\
\rowcolor{PaperStripe}CVE-2023-34095 & 787 & openprinting/cpdb-libs & \texttt{f181bd1f1475} \\
CVE-2023-36274 & 787 & libredwg/libredwg & \texttt{8651fa27dd2d} \\
\rowcolor{PaperStripe}CVE-2023-41910 & 125 & lldpd/lldpd & \texttt{a9aeabdf879c} \\
CVE-2023-47470 & 787 & ffmpeg/ffmpeg & \texttt{4565747056a1} \\
\rowcolor{PaperStripe}CVE-2023-49285 & 125 & squid-cache/squid & \texttt{77b3fb4df0f1} \\
CVE-2023-50268 & 787 & jqlang/jq & \texttt{c9a51565214e} \\
\rowcolor{PaperStripe}CVE-2024-8946 & 787 & micropython/micropython & \texttt{29943546343c} \\
CVE-2024-13903 & 787 & quickjs-ng/quickjs & \texttt{99c02eb45170} \\
\rowcolor{PaperStripe}CVE-2024-25629 & 125 & c-ares/c-ares & \texttt{a804c04ddc82} \\
CVE-2024-32658 & 125 & freerdp/freerdp & \texttt{1a755d898ddc} \\
\rowcolor{PaperStripe}CVE-2024-37407 & 125 & libarchive/libarchive & \texttt{b6a979481b7d} \\
CVE-2024-45306 & 787 & vim/vim & \texttt{396fd1ec2956} \\
\rowcolor{PaperStripe}CVE-2025-9300 & 787 & saitoha/libsixel & \texttt{316c086e79d6} \\
CVE-2025-9394 & 416 & podofo/podofo & \texttt{22d16cb142f2} \\
\rowcolor{PaperStripe}CVE-2025-12875 & 787 & makesoftwaresafe/mruby & \texttt{93619f06dd37} \\
CVE-2025-13654 & 787 & zevv/duc & \texttt{8638c4365ffd} \\
\rowcolor{PaperStripe}CVE-2025-14956 & 125 & webassembly/binaryen & \texttt{4f52bff8c407} \\
CVE-2025-14958 & 787 & seyhajin/sokol & \texttt{33e2271c431b} \\
\rowcolor{PaperStripe}CVE-2025-15155 & 787 & floooh/sokol & \texttt{5d1134415097} \\
CVE-2025-15555 & 787 & open5gs/open5gs & \texttt{54dda0412110} \\
\rowcolor{PaperStripe}CVE-2025-34451 & 787 & httpsgithu/proxychains-ng & \texttt{cc005b713281} \\
CVE-2025-43962 & 125 & libraw/libraw & \texttt{66fe663e02a4} \\
\rowcolor{PaperStripe}CVE-2025-58050 & 787 & pcre2project/pcre2 & \texttt{a141712e5967} \\
CVE-2025-64330 & 787 & oisf/suricata & \texttt{482e5eac9218} \\
\rowcolor{PaperStripe}CVE-2025-64524 & 787 & openprinting/cups-filters & \texttt{956283c74a34} \\
CVE-2025-68973 & 787 & gpg/gnupg & \texttt{115d138ba599} \\
\rowcolor{PaperStripe}CVE-2026-1418 & 787 & enocknt/gpac & \texttt{10c73b82cf0e} \\
CVE-2026-1998 & 787 & dpgeorge/micropython & \texttt{570744d06c5b} \\
\rowcolor{PaperStripe}CVE-2026-2016 & 787 & happyfish100/libfastcommon & \texttt{82f66af3e252} \\
CVE-2026-2258 & 787 & aardappel/lobster & \texttt{c2047a33e1ac} \\
\rowcolor{PaperStripe}CVE-2026-3308 & 190 & artifexsoftware/mupdf & \texttt{a26f0142e7d3} \\
CVE-2026-3945 & 190 & tinyproxy/tinyproxy & \texttt{969852ccdb1d} \\
\rowcolor{PaperStripe}CVE-2026-12630 & 125 & zephyrproject-rtos/zephyr & \texttt{1bbb7aefa69e} \\
CVE-2026-14461 & 125 & traviscross/mtr & \texttt{48e1794414d3} \\
\rowcolor{PaperStripe}CVE-2026-18238 & 125 & the-tcpdump-group/libpcap & \texttt{b9590d482986} \\
CVE-2026-18370 & 787 & eradman/entr & \texttt{2467fe0aa5f2} \\
\rowcolor{PaperStripe}CVE-2026-38821 & 787 & opennds/opennds & \texttt{3b5f7ef40cd0} \\
CVE-2026-39919 & 787 & artifexsoftware/ghostpdl & \texttt{0a8bf88e39db} \\
\rowcolor{PaperStripe}CVE-2026-44235 & 125 & alanxz/rabbitmq-c & \texttt{1d3afbb056fe} \\
CVE-2026-45809 & 787 & opensips/opensips & \texttt{c5970d3ee25b} \\
\rowcolor{PaperStripe}CVE-2026-48977 & 787 & openslide/openslide & \texttt{2be88bd782d9} \\
CVE-2026-50144 & 787 & tencent/ncnn & \texttt{5a0288f255da} \\
\rowcolor{PaperStripe}CVE-2026-50538 & 787 & libvnc/libvncserver & \texttt{540332be3e0a} \\
CVE-2026-53587 & 125 & libgit2/libgit2 & \texttt{2c0ce8c0132a} \\
\rowcolor{PaperStripe}CVE-2026-53938 & 787 & openidc/cjose & \texttt{8c51d2452735} \\
CVE-2026-54616 & 125 & m2team/nanazip & \texttt{733e8570d2ba} \\
\rowcolor{PaperStripe}CVE-2026-55777 & 125 & allinurl/goaccess & \texttt{ba813ed97d99} \\
CVE-2026-56109 & 787 & alsa-project/alsa-lib & \texttt{536dd6f8affd} \\
\rowcolor{PaperStripe}CVE-2026-56684 & 416 & valkey-io/valkey & \texttt{7cd5bcb7575d} \\
CVE-2026-57162 & 121 & pjsip/pjproject & \texttt{a1b707c0c9b0} \\
\rowcolor{PaperStripe}CVE-2026-61548 & 787 & rsyslog/rsyslog & \texttt{bcda60a3692e} \\
CVE-2026-61714 & 125 & fluidsynth/fluidsynth & \texttt{772702e00cc6} \\
\rowcolor{PaperStripe}CVE-2026-62959 & 125 & coturn/coturn & \texttt{960835886692} \\
CVE-2026-63388 & 787 & libevent/libevent & \texttt{52057cb33d0c} \\
\rowcolor{PaperStripe}CVE-2026-65832 & 125 & deskflow/deskflow & \texttt{205a3c803e52} \\
CVE-2026-68768 & 787 & hashcat/hashcat & \texttt{68f56a2d8712} \\
\rowcolor{PaperStripe}CVE-2026-76844 & 787 & madler/zlib & \texttt{df84af25dc19} \\
CVE-2026-77219 & 190 & emacs-mirror/emacs & \texttt{b07e634e4cf4} \\
\rowcolor{PaperStripe}CVE-2026-77358 & 416 & yhirose/cpp-httplib & \texttt{2f986fd5e56e} \\
CVE-2026-81934 & 416 & redis/redis & \texttt{6d088c335d5c} \\
\rowcolor{PaperStripe}CVE-2026-87823 & 190 & luben/zstd-jni & \texttt{46dcb2a54c8b} \\
CVE-2026-102296 & 787 & zoneminder/zoneminder & \texttt{2596e5fb64c0} \\
\end{longtable}

\end{document}